# SECOND-ORDER ADJOINT SENSITIVITY ANALYSIS PROCEDURE (SO-ASAP) FOR COMPUTING EXACTLY AND EFFICIENTLY FIRST- AND SECOND-ORDER SENSITIVITIES IN LARGE-SCALE LINEAR SYSTEMS: II. ILLUSTRATIVE APPLICATION TO A PARADIGM PARTICLE DIFFUSION PROBLEM

Dan G. Cacuci

Department of Mechanical Engineering, University of South Carolina
E-mail: cacuci@cec.sc.edu

Corresponding author:

Department of Mechanical Engineering, University of South Carolina

300 Main Street, Columbia, SC 29208, USA

Email: cacuci@cec.sc.edu; Phone: (919) 909 9624; Submitted to JCP: August 14, 2014

**ABSTRACT**

This work presents an illustrative application of the *second-order adjoint sensitivity analysis procedure* (*SO-ASAP)* to a paradigm neutron diffusion problem, which is sufficiently simple to admit an exact solution, thereby making transparent the mathematical derivations underlying the *SO-ASAP*. The general theory underlying *SO-ASAP* indicates that, for a physical system comprising $N_\alpha$ parameters, the computation of all of the first- and second-order response sensitivities requires $(2N_\alpha +1)$ "large-scale" computations involving correspondingly constructed adjoint systems, which we called *second adjoint sensitivity systems (SASS)*. Very importantly, however, the illustrative application presented in this work shows that the actual number of adjoint computations needed for computing all of the first- and second-order response sensitivities may significantly less than $(2N_\alpha +1)$ per response. For this illustrative problem, four (4) "large-scale" adjoint computations sufficed for the complete and exact computations of all 4 first- and 10 distinct second-order derivatives. Furthermore, the construction and solution of the *SASS* requires very little additional effort beyond the construction of the adjoint sensitivity system needed for computing the first-order sensitivities. Very significantly, only the *sources* on the right-sides of the diffusion (differential) operator needed to be modified; the left-side of the differential equations (and hence the "solver" in large-scale practical applicatons) remained unchanged.

All of the first-order relative response sensitivities to the model parameters have significantly large values, of order unity. Also importantly, most of the second-order relative sensitivities are just as large, and some even up to twice as large as the first-order sensitivities. We show that the second-order sensitivities have the following major impacts on the computed moments of the response distribution: (a) they cause the "expected value of the response" to differ from the "computed nominal value of the response"; (b) their contributions to the response variances and covariances are relatively minor by comparison to the contributions stemming from the first-order response sensitivities; and (c) they *contribute decisively to causing asymmetries in the response distribution*. Indeed, neglecting the second-order sensitivities would nullify the third-order response correlations, and hence would nullify the *skewness* of the response; consequently, any events occurring in a response's long and/or short tails, which are characteristic of rare but decisive events (e.g., major accidents, catastrophes), would likely be missed. We expect the *SO-ASAP* to affect significantly other fields that need efficiently computed second-order response sensitivities, e.g., optimization, data assimilation/adjustment, model calibration, and predictive modeling.

# 1. INTRODUCTION

The accompanying PART I [1] of this work has presented the *second-order forward and adjoint sensitivity analysis procedures* (*SO-FSAP* and *SO-ASAP*) for computing exactly and efficiently the second-order functional derivatives of physical system responses (i.e., "system performance parameters") to the system's model parameters. The definition of "system parameters" includes all computational input data, correlations, initial and/or boundary conditions, etc. The *SO-ASAP* builds on the first-order adjoint sensitivity analysis procedure (*ASAP*) for nonlinear systems introduced in ([2], [3]) and further developed in ([4]-[6]); for recent applications, see also [7] and references therein. For a physical system comprising $N_\alpha$ parameters and $N_r$ responses, we noted in PART I [1] that the *SO-FSAP* requires a total of $\left( N_\alpha^2 / 2 + 3 N_\alpha / 2 \right)$ large-scale computations for obtaining all of the first- and second-order sensitivities, for all $N_r$ system responses. On the other hand, the *SO-ASAP* requires a total of $\left( 2 N_\alpha + 1 \right)$ large-scale computations for obtaining all of the first- and second-order sensitivities, for one functional-type system responses. Therefore, the *SO-FSAP* should be used when $N_r \gg N_\alpha$, while the *SO-ASAP* should be used when $N_\alpha \gg N_r$. The latter case, where the number of system parameters is significantly greater than responses, is the situation most often encountered in practice.

In this work, we present an illustrative application of the *SO-ASAP* to a particle diffusion problem. This paradigm problem comprises the major ingredients needed for highlighting the salient features involved in applying the *SO-ASAP*, yet is sufficiently simple to admit an exact solution, thereby making transparent the mathematical derivations presented in PART I [1]. Very importantly, this application will show that the construction and solution of the second adjoint sensitivity system (*SASS*) requires very little additional effort beyond the construction of the adjoint sensitivity system needed for computing the first-order sensitivities, and that the actual adjoint computations needed for computing all of the first- and second-order response sensitivities are far less than $2 N_\alpha$ per response.

This paper is structured as follows: after defining the particle diffusion problem in Section 2 Section 2.1 and 2.2 present the application of the *SO-ASAP* for obtaining the exact expressions of both the first- and second-order sensitivities of a representative (detector) response to all of model's parameters. In Section 2.3, the numerical values of these sensitivities are used to highlight their fundamental roles in the propagation of model parameter uncertainties to cause uncertainties in the computed responses. Very importantly, it will be shown that they *contribute decisively to causing assymetries in the response distribution*. In other words, neglecting the second-order sensitivities would nullify the third-order response correlations and hence would nullify the response's *skewness*. Consequently, any events occurring in a response's long and/or short tails, which are characteristic of rare but decisive events (e.g., major accidents, catastrophies), would likely be missed. Higher-order moments of the response uncertainty distribution impact subsequent uses of the response distribution, particularly for risk quantification, decision analysis, and quantifying confidence intervals, to name a few. Finally, the concluding remarks in Section 3 highlight directions we are currently pursuing for further generalizing the *SO-ASAP*.

## 2. A PARADIGM PARTICLE DIFFUSION PROBLEM

Consider the diffusion of neutrons in a spent fuel pool like those used by utilities to cool spent fuel reactor elements. For simplicity, the pool is considered to have one direction (say, the length) to be much larger that the other two, so that it could be considered to be a "slab" of water. The fuel elements are considered to be uniformly distributed within the pool, and are considered to emit $Q$ neutrons$/$cm$^3 \cdot$second. The neutrons are considered to be mono-energetic, all possesing the average thermal energy corresponding to the pool's water temperature. Mathematically, this simple physical system can be modeled by using the linear neutron diffusion equation

$$D\frac{d^2\varphi}{dx^2} - \Sigma_a \varphi + Q = 0, \quad x \in (-a, a), \quad \varphi(\pm a) = 0, \tag{1}$$

where $\varphi(x)$ denotes the neutron flux, $D$ denotes the diffusion coefficient, $\Sigma_a$ denotes the macroscopic absorption cross section, $Q$ denotes the distributed source term, and the pool is considered to be a slab of extrapolated thickness $2a$. In view of the problem's symmetry, the origin $x = 0$ can be conveniently chosen at the middle (center) of the slab. For simplicity, the boundary conditions chosen for Eq. (1) are that the neutron flux must vanish at the extrapolated distance $x = a$.

The linear differential equation in Eq. (1) can be solved readily to obtain the general solution

$$\varphi(x) = \frac{Q}{\Sigma_a}\left[1 - \frac{\cosh(xk)}{\cosh(ak)}\right], \quad k \equiv \sqrt{\Sigma_a/D}, \quad x \in (-a, a). \tag{2}$$

where $k \equiv \sqrt{\Sigma_a/D}$ denotes the reciprocal diffusion length. The above expression for the neutron flux $\varphi(x)$ holds for any physical values of $D$, $\Sigma_a$ and $Q$. For practical problems, the flux cannot be obtained analytically as above, but would need to be computed numerically, using the nominal parameter values $\Sigma_a^0$, $D^0$, and $Q^0$. Such a numerical computation would yield the nominal value of the flux, $\varphi^0(x)$, namely

$$\varphi^0(x) = \frac{Q^0}{\Sigma_a^0}\left[1 - \frac{\cosh(xk^0)}{\cosh(ak^0)}\right], \quad k^0 \equiv \sqrt{\Sigma_a^0/D^0}, \quad x \in (-a,a). \tag{3}$$

A generic response, $R(x)$, for the paradigm neutron diffusion problem modeled by Eq. (1) is the reading of a detector within the slab, for example, at a distance $b$ from the slab's midline at $x = 0$. The "reading" of such a detector response would indicate the neutrons' reaction rate within the detectors reactive material, and would be mathematically represented in the form

$$R(\mathbf{e}) \equiv \Sigma_d \varphi(b), \quad \mathbf{e} = (\mathbf{u}, \boldsymbol{\alpha}), \quad \boldsymbol{\alpha} = (\Sigma_a, D, Q, \Sigma_d)^\dagger, \tag{4}$$

where $\Sigma_d$ represents the detector's equivalent reaction cross section, assumed to be constant, for simplicity.

The system parameters for this problem are thus the positive scalars $\Sigma_a$, $D$, $Q$, and $\Sigma_d$. Referring to the mathematical notation used in PART I [1], we note the correspondences

$$\boldsymbol{\alpha} = (\Sigma_a, D, Q, \Sigma_d)^\dagger, \quad \mathbf{u}(\mathbf{x}) \to \varphi(x), \quad \mathbf{e} = \mathbf{e} \equiv (\varphi, \boldsymbol{\alpha})^\dagger, \tag{5}$$

where the dagger $(\dagger)$ denotes transposition. The nominal values of the state function (the neutron flux –in this illustrative problem) and model parameters will be denoted as

$$\boldsymbol{\alpha}^0 = (\Sigma_a^0, D^0, Q^0, \Sigma_d^0)^\dagger, \quad \mathbf{u}^0(\mathbf{x}) \to \varphi^0(x), \quad \mathbf{e}^0 \equiv (\varphi^0, \boldsymbol{\alpha}^0)^\dagger. \tag{6}$$

Using Eq. (3) in Eq. (4) gives the following form for the nominal response value, $R(\mathbf{e}^0)$:

$$R(\mathbf{e}^0) = \Sigma_d^0 \varphi^0(b) = \frac{Q^0 \Sigma_d^0}{\Sigma_a^0}\left[1 - \frac{\cosh(bk^0)}{\cosh(ak^0)}\right], \quad \mathbf{e}^0 \equiv (\varphi^0, \boldsymbol{\alpha}^0)^\dagger, \quad x \in (-a,a). \tag{7}$$

For a fixed value of $b$, the response $R(\mathbf{e})$ is a functional (i.e., a scalar-valued operator) that acts linearly on $\varphi$, as evidenced by in Eq. (7). On the other hand, even though Eq. (1) is linear in $\varphi$, the solution $\varphi(x)$ depends nonlinearly on $\boldsymbol{\alpha}$, as highlighted by Eq. (3). The same is true of the response $R(\mathbf{e})$. Even though $R(\mathbf{e})$ is linear separately in $\varphi$ and in $\boldsymbol{\alpha}$, as shown in Eq. (4), $R(\mathbf{e})$ *is not simultaneously linear in $\varphi$ and $\boldsymbol{\alpha}$*; this fact leads to a nonlinear dependence of $R(\mathbf{e})$ on $\boldsymbol{\alpha}$. This fact is explicit higlighted by the exact expression of $R(\mathbf{e})$ given in Eq. (7).

## *2.1. SO-ASAP Application to Compute Exactly and Efficiently the First-Order Response Sensitivities to Model Parameters*

The parameters in this problem are determined from experiments afflicted by uncertainties; in particular, the uncertainties for the basic (microscopic) neutron cross sections are provided in centrally deposited "covariance files". Therefore, the system parameters $\boldsymbol{\alpha}$ can an do vary (because of uncertainties, external influences, etc.) from their nominal values $\boldsymbol{\alpha}^0$ by amounts denoted here as

$$\mathbf{h}_\alpha \equiv \left( \delta\Sigma_a, \delta D, \delta Q, \delta\Sigma_d \right)^\dagger, \tag{8}$$

For easy reference, the notation in Eq. (8) corresponds to the notation used in PART I [1]. In practice, the variations $\delta\Sigma_a, \delta D, \delta Q, \delta\Sigma_d$, usually correspond to the standard deviations quantifying the uncertainties in the respective model parameters. As shown in PART I [1], the sensitivities of the response $R(\mathbf{e})$ to the variations $\mathbf{h}_\alpha$ are given by the (first-order) G-differential $\delta R(\mathbf{e}^0; \mathbf{h})$ of $R(\mathbf{e})$ at $\mathbf{e}^0 \equiv (\boldsymbol{\alpha}^0, \varphi^0)$, which is defined as

$$\delta R(\mathbf{e}^0; \mathbf{h}) \equiv \frac{d}{d\varepsilon} \left\{ R(\mathbf{e}^0 + \varepsilon \mathbf{h}) \right\}_{\varepsilon=0}, \text{ with } \mathbf{h} \equiv (h_\varphi, \mathbf{h}_\alpha)^\dagger. \tag{9}$$

Applying the above definition to the response defined by Eq. (3) gives

$$\delta R\left(\mathbf{e}^{0};\mathbf{h}\right) \equiv \frac{d}{d\varepsilon}\left\{\left[\Sigma_{d}^{0}+\varepsilon\left(\delta\Sigma_{d}\right)\right]\left[\varphi^{0}(x)+\varepsilon\, h_{\varphi}(x)\right]\right\}_{\varepsilon=0} = \mathbf{R}'_{\alpha}\left(\mathbf{e}^{0}\right)\mathbf{h}_{\alpha} + R'_{\varphi}\left(\mathbf{e}^{0}\right)h_{\varphi}, \quad (10)$$

where the "direct-effect" term $\mathbf{R}'_{\alpha}\left(e^{0}\right)\mathbf{h}_{\alpha}$ is defined as

$$\mathbf{R}'_{\alpha}\left(e^{0}\right)\mathbf{h}_{\alpha} \equiv \varphi^{0}(b)\left[(0,0,0,1)\cdot\left(\delta\Sigma_{a},\delta D,\delta Q,\delta\Sigma_{d}\right)^{\dagger}\right] = \left(\delta\Sigma_{d}\right)\varphi^{0}(b), \quad (11)$$

while the "indirect-effect" term $R'_{\varphi} h_{\varphi}$ is defined as

$$R'_{\varphi}\left(\mathbf{e}^{0}\right)h_{\varphi} \equiv \Sigma_{d}^{0} h_{\varphi}(x). \quad (12)$$

As Eq. (10) indicates, the operator $\delta R\left(\mathbf{e}^{0};\mathbf{h}\right)$ is linear in $\mathbf{h}$; in particular, $R'_{\varphi} h_{\varphi}$ is a linear operator on $h_{\varphi}$. This linear dependence on $\mathbf{h}$ is underscored by writing henceforth $DR\left(\mathbf{e}^{0};\mathbf{h}\right)$ instead of $\delta R\left(\mathbf{e}^{0};\mathbf{h}\right)$ to denote the sensitivity of $R(\mathbf{e})$ at $\mathbf{e}^{0}$ to variations $\mathbf{h}$.

The "direct-effect" term $\mathbf{R}'_{\alpha}\left(e^{0}\right)\mathbf{h}_{\alpha}$ can be evaluated at this stage by replacing Eq. (6) into Eq. (11), to obtain

$$\mathbf{R}'_{\alpha}\left(e^{0}\right)\mathbf{h}_{\alpha} = \left(\delta\Sigma_{d}\right)\frac{Q^{0}}{\Sigma_{a}^{0}}\left[1-\frac{\cosh\left(xk^{0}\right)}{\cosh\left(ak^{0}\right)}\right], \quad x\in(-a,a). \quad (13)$$

However, the "indirect-effect" term, $R'_{\varphi} h_{\varphi}$, cannot be evaluated at this stage, since $h_{\varphi}(x)$ is not yet available. As described in PART I [1], $h_{\varphi}(x)$ is related to $\mathbf{h}_{\alpha}$, and this relationship is given by the "forward sensitivity equations" (FSE) obtained by computing the G-differentials of Eq. (1), together with the corresponding boundary conditions. Performing this operation gives the equation

$$L\left(\boldsymbol{\alpha}^{0}\right)h_{\varphi} + \left[L'_{\alpha}\left(\boldsymbol{\alpha}^{0}\right)\varphi^{0}\right]\mathbf{h}_{\alpha} = 0, \quad (14)$$

together with the boundary conditions

$$h_\varphi(\pm a) = 0. \tag{15}$$

In Eq. (14), the operator $L(\boldsymbol{\alpha}^0)$ is defined as

$$L(\boldsymbol{\alpha}^0) \equiv D^0 \frac{d^2}{dx^2} - \Sigma_a^0, \tag{16}$$

while the operator $\left[L'_\alpha(\boldsymbol{\alpha}^0)\varphi^0\right]\mathbf{h}_\alpha$ is defined below:

$$\left[L'_\alpha(\boldsymbol{\alpha}^0)\varphi^0\right]\mathbf{h}_\alpha \equiv (\delta D)\frac{d^2\varphi^0}{dx^2} - (\delta\Sigma_a)\varphi^0 + (\delta Q), \tag{17}$$

Thus, the operator $\left[L'_\alpha(\boldsymbol{\alpha}^0)\varphi^0\right]\mathbf{h}_\alpha$, which mathematically represents the partial G-differential of $(L\varphi)$ at $\boldsymbol{\alpha}^0$ with respect to $\boldsymbol{\alpha}$, contains all of the first-order parameter variations $\mathbf{h}_\alpha$.

Equation (14) together with the boundary conditions in Eq. (15) constitute the "*forward sensitivity system*". To obtain its solution, $h_\varphi(x)$, for every possible variation $\mathbf{h}_\alpha$, this system of equations would need to be solved repeatedly (each such computation represents a "large-scale" computation). Thus, the *forward sensitivity system* is advantageous to use only if the number of responses exceeds the number of model parameters, which is seldom the case in practice. Nevertheless, Eqs. (14) and (15) provide an independent way to verify the results that will be obtained later in this Section by applying the *SO-ASAP*. Therefore, we provide below the expression of the solution of Eqs. (14) and (15):

$$\begin{aligned}h_\varphi(x) = &C_1\left[\cosh(xk^0) - \cosh(ak^0)\right] \\ &+ C_2\left[x\sinh(xk^0)\cosh(ak^0) - a\sinh(ak^0)\cosh(xk^0)\right], \ x \in (-a, a).\end{aligned} \tag{18}$$

where the constants $C_1$ and $C_2$ are defined, respectively, as

$$C_1 \equiv \frac{\left[(\delta \Sigma_a)Q^0/\Sigma_a^0 - (\delta Q)\right]}{\Sigma_a^0 \left(\cosh ak^0\right)} \ , \tag{19}$$

and

$$C_2 \equiv \frac{\left[(\delta D)/D^0 - (\delta \Sigma_a)/\Sigma_a^0\right]Q^0}{2\sqrt{D^0 \Sigma_a^0} \left(\cosh ak^0\right)^2} \ . \tag{20}$$

Hence, the expression of the "indirect-effect" term is obtained by replacing Eq. (18) in Eq. (12) to obtain:

$$\begin{aligned} R'_\varphi \left(\mathbf{e}^0\right) h_\varphi (x) &= \Sigma_d^0 C_1 \left[\cosh\left(xk^0\right) - \cosh\left(ak^0\right)\right] \\ &+ \Sigma_d^0 C_2 \left[x \sinh\left(xk^0\right)\cosh\left(ak^0\right) - a \sinh\left(ak^0\right)\cosh\left(xk^0\right)\right], \ x \in (-a, a). \end{aligned} \tag{21}$$

As was generally shown in [2], and recalling the notations introduced in PART I [1], the fundamental prerequisite for applying the *SO-ASAP* is the introduction of a Hilbert space, $\mathcal{H}_u$, appropriate to the problem at hand. For our illustrative example, it is appropriate to chose $\mathcal{H}_u$ to be the real Hilbert space $\mathcal{H}_u \equiv \mathcal{L}_2(\Omega)$, with $\Omega \equiv (-a, a)$, equipped with the inner product

$$\langle f(x), g(x) \rangle \equiv \int_{-a}^{a} f(x)g(x)dx, \ \text{for} \ f, g \in \mathcal{H}_u \equiv \mathcal{L}_2(\Omega), \ \Omega \equiv (-a, a). \tag{22}$$

In view of Eq. (22), the "indirect-effect" term $R'_\varphi \left(\mathbf{e}^0\right) h_\varphi$ can be expressed in inner product form as follows:

$$R'_\varphi \left(\mathbf{e}^0\right) h_\varphi \equiv \Sigma_d^0 h_\varphi (b) = \int_{-a}^{a} \Sigma_d^0 h_\varphi (x) \delta(x-b) dx = \left\langle \Sigma_d^0 \delta(x-b), h_\varphi \right\rangle. \tag{23}$$

The next step underlying the *SO-ASAP* is the construction of the operator $L^+\left(\boldsymbol{\alpha}^0\right)$ that is formally adjoint to $L\left(\boldsymbol{\alpha}^0\right)$. For this purpose, Eq. (14) is multiplied by a square-integrable, but at this stage otherwise still arbitrary, function $\psi(x) \in \mathcal{H}_Q = \mathcal{L}_2(\Omega)$, and is subsequently integrated over $x$ to obtain

$$\int_{-a}^{a} \psi(x) \left[ D^0 \frac{d^2 h_\varphi}{dx^2} - \Sigma_a^0 h_\varphi(x) \right] dx = -\int_{-a}^{a} \psi(x) \left[ (\delta D) \frac{d^2 \varphi^0}{dx^2} - (\delta \Sigma_a) \varphi^0 + (\delta Q) \right] dx. \quad (24)$$

Integrating the left-side of Eq. (24) by parts twice, to transfer all operations on $h_\varphi(x)$ to operations on $\psi(x)$, leads to

$$\int_{-a}^{a} \psi(x) \left[ D^0 \frac{d^2 h_\varphi}{dx^2} - \Sigma_a^0 h_\varphi(x) \right] dx = \int_{-a}^{a} \left[ D^0 \frac{d^2 \psi(x)}{dx^2} - \Sigma_a^0 \psi(x) \right] h_\varphi(x) dx$$

$$+ D^0 \left[ \psi \frac{dh_\varphi}{dx} - h_\varphi \frac{d\psi}{dx} \right]_{-a}^{a}. \quad (25)$$

The right-side of the above equations shows that the formal adjoint of $L(\boldsymbol{\alpha}^0)$ is the operator

$$L^*(\boldsymbol{\alpha}^0) \equiv D^0 \frac{d^2}{dx^2} - \Sigma_a^0. \quad (26)$$

Note that the function $\psi(x)$ is still arbitrary at this stage, except for the requirement that $\psi \in \mathcal{H}_Q = \mathcal{L}_2(\Omega)$. The next step in the construction of the adjoint system is the identification of the source term, which is achieved by requiring the first terms on the right-sides of Eqs. (23) and (25) to represent the same functional. Imposing this requirement yields the equation

$$L^*(\boldsymbol{\alpha}^0) \psi \equiv D^0 \frac{d^2 \psi}{dx^2} - \Sigma_a^0 \psi(x) = \Sigma_d^0 \delta(x - b), \quad (27)$$

The boundary conditions for $\psi(x)$ can now be selected by requiring that unknown values of $h_\varphi$, such as the derivatives $\{dh_\varphi/dx\}_{-a}^{a}$, would be eliminated from Eq. (25). Since $h_\varphi$ is known at $x = \pm a$ from Eq. (15), the elimination of unknown values of $h_\varphi$ can be accomplished by choosing the boundary conditions

$$\psi(\pm a) = 0. \qquad (28)$$

The above selection of the adjoint boundary conditions completes the construction of the *adjoint sensitivity system*, consisting of Eqs. (27) and (28). Since the boundary conditions in Eq. (28) for the adjoint function $\psi(x)$ are the same as the boundary conditions for $h_\phi(x)$ in Eq. (15), it follows from this and from Eqs. (16) and (26) that the operators $L^*(\boldsymbol{\alpha}^0)$ and $L(\boldsymbol{\alpha}^0)$ are not just formally, but bona-fide self-adjoint.

Using Eqs.(24) and (27) in Eq.(23) leads to the following expression for the "indirect-effect" term $R'_\varphi(\mathbf{e}^0)h_\varphi$ in terms of the adjoint function $\psi(x)$:

$$R'_\varphi(\mathbf{e}^0)h_\varphi = -\int_{-a}^{a}\psi(x)\left[(\delta D)\frac{d^2\varphi^0}{dx^2} - (\delta \Sigma_a)\varphi^0(x) + (\delta Q)\right]dx. \qquad (29)$$

As expected, *the adjoint sensitivity system,* cf., Eqs. (27) and (28), *is independent of parameter variations* $\mathbf{h}_\alpha$, so it needs to be solved only once to obtain the adjoint function $\psi(x)$. Very important, too, is the fact (characteristic of linear systems) that the adjoint system is independent of the original solution $\varphi^0(x)$ and can therefore be solved independently (and without any knowledge) of the forward neutron flux $\varphi(x)$. Of course, the adjoint system depends on the response, which provides the source term as shown in Eq. (27). Solving the adjoint system for our illustrative example yields the following expression for the adjoint function $\psi(x)$:

$$\psi(x) = \frac{\Sigma_d^0}{\sqrt{\Sigma_a^0 D^0}}\left\{\frac{\sinh\left[(b-a)k^0\right]}{\sinh(2ak^0)}\sinh\left[(x+a)k^0\right] + H(x-b)\sinh\left[(x-b)k^0\right]\right\}, \qquad (30)$$

where $H(x-b)$ is the Heaviside-step functional defined as

$$H(x) = \begin{cases} 0, & \text{for } x < 0 \\ 1, & \text{for } x \geq 0 \end{cases}. \tag{31}$$

Replacing the adjoint function $\psi(x)$ from Eq. (30) in Eq. (29) and carrying out the respective integration over $x$ yields the first-order sensitivity $DR(\mathbf{e}^0;\mathbf{h})$ of $R(\mathbf{e})$ at $\mathbf{e}^0$ to variations $\mathbf{h}_\alpha$ in the system parameters:

$$DR(\mathbf{e}^0;\mathbf{h}) = \frac{\partial R}{\partial \Sigma_a}(\delta \Sigma_a) + \frac{\partial R}{\partial D}(\delta D) + \frac{\partial R}{\partial Q}(\delta Q) + \frac{\partial R}{\partial \Sigma_d}(\delta \Sigma_d), \tag{32}$$

where

$$S_1(\alpha^0) \triangleq \frac{\partial R}{\partial \Sigma_a} = \int_{-a}^{a} \psi^0(x) \varphi^0(x) \, dx$$
$$= Q^0 \Sigma_d^0 \left[ -\left(\Sigma_a^0\right)^{-2} A(k^0) + \frac{1}{2\Sigma_a^0 \sqrt{D^0 \Sigma_a^0}} B(k^0) \right] \tag{33}$$

$$S_2(\alpha^0) \triangleq \frac{\partial R}{\partial D} = \int_{-a}^{a} \psi^0(x) \frac{d^2 \varphi^0}{dx^2} \, dx = -\frac{\Sigma_a^0}{D^0} \int_{-a}^{a} \psi^0(x) \varphi^0(x) \, dx + \frac{Q^0}{D^0} \int_{-a}^{a} \psi^0(x) \, dx$$
$$= -\frac{Q^0 \Sigma_d^0}{2D^0 \Sigma_a^0} k^0 B(k^0) \tag{34}$$

$$S_3(\alpha^0) \triangleq \frac{\partial R}{\partial Q} = -\int_{-a}^{a} \psi^0(x) \, dx = \Sigma_d^0 \left(\Sigma_a^0\right)^{-1} A(k^0) \tag{35}$$

$$S_4(\alpha^0) \triangleq \frac{\partial R}{\partial \Sigma_d} = \int_{-a}^{a} \varphi^0(x) \delta(x-b) \, dx = Q^0 \left(\Sigma_a^0\right)^{-1} A(k^0) \tag{36}$$

with the quatities $A(k^0)$ and $B(k^0)$ defined, respectively, as

$$A(k^0) \triangleq 1 - \frac{\cosh(bk^0)}{\cosh(ak^0)}, \tag{37.a}$$

$$B(k^0) \triangleq \left.\frac{dA}{dk}\right|_{k^0} = \left[\cosh(ak^0)\right]^{-2} \left[a\sinh(ak^0)\cosh(bk^0) - b\sinh(bk^0)\cosh(ak^0)\right].$$

(37.b)

One of the main uses of sensitivities is for ranking the relative importance of parameter variations in influencing variations in responses. Relative sensitivities are used for this purpose, since they are dimensionless numbers; the relative sensitivity of a response $R(\mathbf{e})$ to the $i^{th}$-parameter, $\alpha_i$, is defined as $S_i^{rel} \triangleq (\partial R/\partial \alpha_i)_{\mathbf{e}^0} \left[\alpha_i^0 / R(\mathbf{e}^0)\right]$. It is evident from Eqs. (7), (35) and (36) that the relative sensitivities of the detector response $R(\mathbf{e})$ to the source $(\alpha_3 \triangleq Q)$ and to the detector's reaction cross section $(\alpha_4 \triangleq \Sigma_d)$ are unity, i.e., $S_3^{rel} \triangleq (\partial R/\partial Q)_{\mathbf{e}^0} \left[Q^0 / R(\mathbf{e}^0)\right] = 1$ and $S_4^{rel} \triangleq (\partial R/\partial \Sigma_d)_{\mathbf{e}^0} \left[\Sigma_d^0 / R(\mathbf{e}^0)\right] = 1$, regardless of the position $(x = b)$ of the location of the detector or the material properties of the medium. This means that a *1%* increase/decrease in either $Q$ or $\Sigma_d$ will induce a *1%* increase/decrease in nominal value of the response $R(\mathbf{e})$. These are rather large sensitivities, indeed! The relative influence of a response's sensitivities to the medium's macroscopic absorption cross section, $\Sigma_a$, and to the medium's diffusion coefficient, $D$, can best be illustrated by considering, specifically, that $a = 50\ cm$, and the distributed neutron sources emit nominally $S^0 = 10^7\ neutrons \cdot cm^{-3} \cdot s^{-1}$. Furthermore, the material properties describing the absorption and diffusion of thermal neutron in water have the following nominal values: $\Sigma_a^0 = 0.0197\ cm^{-1}$, $D^0 = 0{,}16\ cm$. Consider that measurements are performed with an idealized, infinitely thin, detector having an indium-like nominal detector cross section $\Sigma_d^0 = 7.438\ cm^{-1}$.

Consider now six symmetric responses, at the following detector locations: (i) $b = \pm 10\ cm$, located close to the water pool's center; (ii) $b = \pm 40\ cm$, located close to the pool's boundaries; and (iii) $b = \pm 49.5\ cm$, located very close to the edge, where the neutron flux' gradient becomes very steep, and the diffusion coefficient markedly influences the detector's response. We designate the detector's responses at these locations as *R₁ (10 cm), R₂ (40 cm), R₃ (49.5 cm), R₄ (-10 cm), R₅ (-40 cm),* and *R₆ (-49.5 cm)*. These symmetric locations, at which Eq. (7) clearly indicates that $R(x) = R(-x)$, were deliberately chosen in order to verify the preservation of symmetry during and after the computation of the respective responses' sensitivities to the model parameters. In units of $[neutrons \cdot cm^{-3} \cdot s^{-1}]$, the nominal computed response values obtained using Eq. (7) for these detector locations were as follows:

$R_1(10\,cm) = R_4(-10\,cm) = 3.77 \times 10^9$, $\quad R_2(40\,cm) = R_4(-40\,cm) = 3.66 \times 10^9$, and $R_3(49.5\,cm) = R_6(-49.5\,cm) = 6.076 \times 10^8$.

Note also that $R(x) = R(-x)$, so that only *three* (rather than six) *"large-scale" adjoint sensitivity computations* [i.e., soving Eqs. (27) and (28)] suffice to obtain all of results presented in Table 1 (absolute sensitivities, in the respective units, which are, for brevity omitted from the table) and Table 2 (relative sensitivities).

Table 1. Absolute sensitivities for six detector responses

| First-order Absolute Sensitivities | $R_1$ (10 cm) $R_4$ (-10 cm) | $R_2$ (40 cm) $R_5$ (-40 cm) | $R_3$ (49.5 cm) $R_6$ (-49.5 cm) |
|---|---|---|---|
| $S_1$ (SIGa) | -1.917x10¹¹ | -1758x10¹¹ | -1.673x10¹⁰ |
| $S_2$ (D) | -1.331x10⁵ | -1.239x10⁹ | -1.737x10⁹ |
| $S_3$ (Q) | 3.776x10² | 3.663x10² | 6.076x10¹ |
| $S_4$ (SIGd) | 5.076x10⁸ | 4.924x10⁸ | 8.168x10⁷ |

Table 2. Relative sensitivities for six detector responses

| First-order Relative Sensitivities | $R_1$ (10 cm) $R_4$ (-10 cm) | $R_2$ (40 cm) $R_5$ (-40 cm) | $R_3$ (49.5 cm) $R_6$ (-49.5 cm) |
|---|---|---|---|
| $S_1$ (SIGa)_rel | -1.00 | -0.95 | -0.54 |
| $S_2$ (D)_rel | -5.64x10⁻⁶ | -0.05 | -0.46 |
| $S_3$ (Q)_rel | 1.00 | 1.00 | 1.00 |
| $S_4$ (SIGd)_rel | 1.00 | 1.00 | 1.00 |

The absolute sensitivities will be used in Section 2.3 to compute response uncertainties arising from parameter uncertainties. They are difficult to use, however, for ranking the importance of parameters in influencing the respective responses. By contrast, the relative sensitivities presented in Table 2 readily indicate the importance of the model parameters in influencing the detector responses. Of course, these "parameter importances" reflect the relative importance of the physical processes associated with the respective parameters in influencing the detector's response. As already mentioned, $S_3^{rel} = S_4^{rel} = 1$, and these sensitivities turn out to be the most important ones for the detector response, regardless of the detector's position. This result is physically due to the fact that the neutron source strength, on the one hand, and the detector's interaction cross section, should be very important. Note, in particular, that the sensitivity $S_4^{rel}$ describes a "direct effect term" specific to the detector's properties (in this illustrative paradigm example, $S_4^{rel}$ is actually independent of the medium's material properties).

The other two relative sensitivities, namely $S_1^{rel} \triangleq (\partial R/\partial \Sigma_a)_{\mathbf{e}^0} \left[ \Sigma_a^0 / R(\mathbf{e}^0) \right]$ and $S_2^{rel} \triangleq (\partial R/\partial D)_{\mathbf{e}^0} \left[ D^0 / R(\mathbf{e}^0) \right]$ describe opposing physical processes and, hence, influences on the detector responses. Thus, $S_1^{rel}$ is very important in the center of the slab, where it has a large negative value of comparable absolute magnitude to $S_3^{rel}$. Towards the edge of the slab, $S_1^{rel}$ decreases in absolute magnitude but remains negative throughout the medium. Physically, the behavior of $S_1^{rel}$ indicates that the absorption process of neutrons cross section is a loss mechanism (hence the negative sign of $S_1^{rel}$ throughout the medium), and its importance is almost as large as the production by the neutron sources (as indicated by the positive sign and magnitude of $S_3^{rel}$), throughout most of the medium. Very close to the physical boundary of the medium, where the neutron flux must vanish because of the chosen boundary conditions, the absorption mechanism diminishes rapidly in importance. The flux gradient is very steep over a small distance towards the boundary, indicating a boundary-layer-like behavior of the absorption process, which obviously must cease at the medium's extrapolated boundary, where the flux must mathematically vanish due to the imposed boundary conditions.

On the other hand, the behavior of $S_2^{rel} \triangleq (\partial R/\partial D)_{\mathbf{e}^0} \left[ D^0 / R(\mathbf{e}^0) \right]$ indicates that the diffusion coefficient and consequently the details of the diffusion process, play a very minor role in influencing the detector response, except close to the boundaries of the slab. The sign of $S_2^{rel}$ is negative, indicating a "loss" of neutrons in contributing to the detector's response (i.e., increasing the diffusion coefficient, and hence the importance of the diffusion process, causes fewer neutrons to interact with the detector and hence contribute to its response). Towards the extrapolated boundary of the medium, namely at 0.5 cm from the respective extrapolated boundaries, the magnitude of $S_2^{rel}$ becomes comparable to that of $S_1^{rel}$, both being negative, indicating neutron "losses" from the detector's response. This result is expected since the neutron flux was (mathematically) required to vanish at the extrapolated boundaries, and the physical mechanisms that reduce the neutron flux can only stem from absoption combined with diffusion.

## 2.2. SO-ASAP Application to Compute Exactly and Efficiently the Second-Order Response Sensitivities to Model Parameters

As shown in the general theory presented in Section 3.1 of PART I [1], the fundamental philosophical and starting point for computing the second-order derivatives (sensitivities) $S_{ij}(\alpha^0) \triangleq \partial^2 R/\partial \alpha_i \partial \alpha_j = \partial^2 R/\partial \alpha_j \partial \alpha_i$ was the consideration of the first-order sensitivities to be functionals of the form $S_i(\varphi, \psi, \boldsymbol{\alpha})$, $i = 1, 2, 3, 4$. Based on this fundamental consideration, the *SO-ASAP* proceeds by computing the first-order G-differential, $\delta S_i(\mathbf{e}^0, \boldsymbol{\psi}^0, \mathbf{g})$, of (any of) the functionals $S_i(\mathbf{u}, \boldsymbol{\alpha}, \boldsymbol{\psi})$, at the point $(\mathbf{e}^0, \boldsymbol{\psi}^0)$, from the definition

$$\delta S_i\left(\varphi^0, \psi^0, \boldsymbol{\alpha}^0; h_\varphi, h_\psi, \mathbf{h}_\alpha\right) \equiv \left\{\frac{d}{d\varepsilon}\left[S_i\left(\varphi^0 + \varepsilon h_\varphi, \psi^0 + \varepsilon h_\psi, \boldsymbol{\alpha}^0 + \varepsilon \mathbf{h}_\alpha\right)\right]\right\}_{\varepsilon=0} \tag{38}$$

for an arbitrary scalar $\varepsilon \in \mathcal{F}$, and all (i.e., arbitrary) vectors $(h_\varphi, h_\psi, \mathbf{h}_\alpha) \in \mathcal{H}_\varphi(\Omega_x) \times \mathcal{H}_\psi(\Omega_x) \times \mathcal{H}_\alpha$. For our illustrative example, there are only $N_\alpha(N_\alpha + 1)/2 = 10$ distinct second-order derivatives, due to the symmetry property $\partial^2 R/\partial \alpha_i \partial \alpha_j = \partial^2 R/\partial \alpha_j \partial \alpha_i$. It is convenient to compute the sensitivities $S_{ij}(\alpha^0)$ by starting with the simplest ones and progressing to the more difficult ones. An examination of Eqs. (33) through (36) readily reveals that $S_4(\alpha^0)$ and $S_3(\alpha^0)$ have the simplest expressions, while $S_2(\alpha^0)$ has the most "complicated" expression. Therefore, it is computationally advantageous to compute first the four second-order derivatives $S_{4i}(\alpha^0) \equiv \partial^2 R/\partial Q \partial \alpha_i$, $(1 = 1, 2, 3, 4)$, followed by the three second-order derivatives $S_{3i}(\alpha^0) \equiv \partial^2 R/\partial \Sigma_d \partial \alpha_i$, $(1 = 1, 2, 3)$, since $S_{34}(\alpha^0)$ would already be available. The computations would then proceed by determining $S_{1i}(\alpha^0) \equiv \partial^2 R/\partial \Sigma_a \partial \alpha_i$, $(1 = 1, 2)$ and, lastly, $S_{22}(\alpha^0) \equiv \partial^2 R/\partial D^2$. These computations will be performed explicitly in the following sub-sections.

*2.2.1. Applying the SO-ASAP to Compute the Second-Order Response Sensitivities $S_{4i} \triangleq \partial^2 R/(\partial \Sigma_d \partial \alpha_i)$*

Applying the definition shown in Eq. (38) to Eq. (36) yields the corresponding G-differential, $DS_4(\alpha^0)$, of the first-order sensitivity $S_4(\alpha)$, as

$$DS_4(\alpha^0) = \frac{d}{d\varepsilon}\left\{ \int_{-a}^{a}(\varphi^0 + \varepsilon h_\varphi)\delta(x-b)dx \right\}_{\varepsilon=0} = \int_{-a}^{a} h_\varphi(x)\delta(x-b)dx, \qquad (39)$$

where the function $h_\varphi$ is the solution of Eqs. (14) and (15). Note that the entire contribution to $DS_4(\alpha^0)$ comes from the "indirect-effect" term; there is no "direct-effect" term contribution to $DS_4(\alpha^0)$.

Applying the general theoretical consideration presented in PART I [1] for the *SO-ASAP*, we note that the system adjoint to Eqs. (14) and (15), but corresponding to the "response" $DS_4(\alpha)$, can be obtained by following the procedure outlined in Section 2.1, above, when deriving the (first) *adjoint sensitivity system* for the adjoint function $\psi(x)$, cf., Eqs. (27) and (28). We will designate "the system adjoint to Eqs. (14) and (15), but corresponding to the response $DS_4(\alpha)$" as the *second adjoint sensitivity system, and its solution will be designated as* $\lambda_4(x)$. Thus, following the logic dscribes in Section 2.1, above, we readily obtain the following *second adjoint sensitivity system* for the adjoint function $\lambda_4(x)$:

$$D^0 \frac{d^2 \lambda_4}{dx^2} - \Sigma_a^0 \lambda_4(x) = \delta(x-b), \qquad (40)$$

$$\lambda_4(\pm a) = 0. \tag{41}$$

Since $DS_4(\alpha^0)$ does not depend on $h_\psi(x)$, the second adjoint function [say, $\theta_4(x)$], which would have corresponded to $h_\psi(x)$, is identically zero, since the adjoint system for $\theta_4(x)$ would have been a linear homogeneous equation with zero source and zero boundary conditions. In terms of the adjoint function $\lambda_4(x)$, the expression of the differential $DS_4(\alpha^0)$ becomes:

$$\begin{aligned}DS_4(\alpha^0) &= \int_{-a}^{a} \lambda_4(x) \left[ -(\delta D)\frac{d^2\varphi^0}{dx^2} + (\delta\Sigma_a)\varphi^0(x) - (\delta Q) \right] dx \\ &= S_{41}(\delta\Sigma_a) + S_{42}(\delta D) + S_{43}(\delta Q) + S_{44}(\delta\Sigma_d),\end{aligned} \tag{42}$$

where

$$S_{41} \triangleq \frac{\partial^2 R}{\partial \Sigma_d \partial \Sigma_a} = \int_{-a}^{a} \lambda_4(x)\varphi^0(x) dx, \tag{43}$$

$$S_{42} \triangleq \frac{\partial^2 R}{\partial \Sigma_d \partial D} = -\int_{-a}^{a} \lambda_4(x) \frac{d^2\varphi^0}{dx^2} dx, \tag{44}$$

$$S_{43} \triangleq \frac{\partial^2 R}{\partial \Sigma_d \partial Q} = -\int_{-a}^{a} \lambda_4(x) dx, \tag{45}$$

$$S_{44} \triangleq \frac{\partial^2 R}{\partial \Sigma_d \partial \Sigma_d} \equiv 0. \tag{46}$$

Note that $S_{44} \equiv 0$, since the expression of $DS_4(\alpha^0)$, namely Eq.(39), does not contain the quantity $(\delta\Sigma_d)$. Note also that the adjoint system satisfied by $\lambda_4(x)$, comprising Eqs.(40) and (41), is the same as would be obtained by setting $\Sigma_d^0 \equiv 1$ into the adjoint system satisfied by the adjoint function $\psi(x)$, cf. Eqs. (27) and (28). Therefore, $\lambda_4(x)$ is simply obtained by dividing the result in Eq. (30) by $\Sigma_d^0$ or, equivalently, setting $\Sigma_d^0 \equiv 1$ in Eq. (30), to obtain:

$$\lambda_4(x) = \{\psi(x)\}_{\Sigma_d^0 \equiv 1}$$
$$= \frac{1}{\sqrt{\Sigma_a^0 D^0}} \left\{ \frac{\sinh\left[(b-a)k^0\right]}{\sinh(2ak^0)} \sinh\left[(x+a)k^0\right] + H(x-b)\sinh\left[(x-b)k^0\right] \right\}. \tag{47}$$

Even more, since $\lambda_4(x) = \{\psi(x)\}_{\Sigma_d^0 \equiv 1}$, it is not even necessary to solve Eqs. (40) and (41) to compute $\lambda_4(x)$, and it is not necessary to introduce Eq. (47) into Eqs. (43) through (45), and to perform the respective integrations. In fact, *one simply replaces $\Sigma_d^0 \equiv 1$ in Eqs. (33) through (35), or, equivalently, divides these equations through $\Sigma_d^0$, to obtain:*

$$S_{41}(\alpha^0) \triangleq \frac{\partial^2 R}{\partial \Sigma_d \partial \Sigma_a} = \frac{S_1(\alpha^0)}{\Sigma_d^0} = Q^0 \left[ -\left(\Sigma_a^0\right)^{-2} A(k^0) + \frac{1}{2\Sigma_a^0 \sqrt{D^0 \Sigma_a^0}} B(k^0) \right], \tag{48}$$

$$S_{42}(\alpha^0) \triangleq \frac{\partial^2 R}{\partial \Sigma_d \partial D} = \frac{S_2(\alpha^0)}{\Sigma_d^0} = -\frac{Q^0}{2D^0 \Sigma_a^0} k^0 B(k^0), \tag{49}$$

$$S_{43}(\alpha^0) \triangleq \frac{\partial^2 R}{\partial \Sigma_d \partial Q} = \frac{S_3(\alpha^0)}{\Sigma_d^0} = \left(\Sigma_a^0\right)^{-1} A(k^0). \tag{50}$$

*2.2.2. Applying the SO-ASAP to Compute the Second-Order Response Sensitivities $S_{3i} \triangleq \partial^2 R / (\partial Q \partial \alpha_i)$.*

The G-differential, $DS_3(\alpha^0)$, of the first-order sensitivity $S_3(\alpha)$ is computed by applying the definition given in Eq. (38) to Eq. (35) to obtain

$$DS_3(\alpha^0) = -\frac{d}{d\varepsilon} \left\{ \int_{-a}^{a} \left( \psi^0 + \varepsilon h_\psi \right) dx \right\}_{\varepsilon=0} = -\int_{-a}^{a} h_\psi(x) \, dx, \tag{51}$$

where the function $h_\psi$ is the solution of the G-differentiated (first) adjoint sensitivity system, cf., Eqs. (27) and (28). Thus, applying the definition of the G-differential [cf., Eq. (38)] to Eqs. (27) and (28) yields:

$$D^0 \frac{d^2 h_\psi}{dx^2} - \Sigma_a^0 h_\psi(x) = -(\delta D)\frac{d^2 \psi^0}{dx^2} + (\delta \Sigma_a)\psi^0 + (\delta \Sigma_d^0)\delta(x-b), \qquad (52)$$

$$h_\psi(\pm a) = 0. \qquad (53)$$

Once again, the entire contribution to $DS_3(\alpha^0)$ comes from the "indirect-effect" term; there is no "direct-effect" term contribution to $DS_3(\alpha^0)$. The system adjoint to Eqs. (52) and (53), and also corresponding to the "response" $DS_3(\alpha)$ is constructed by applying the same procedure as already used in the previous Sections, to obtain:

$$D^0 \frac{d^2 \theta_3}{dx^2} - \Sigma_a^0 \theta_3(x) = 1, \qquad (54)$$

$$\theta_3(\pm a) = 0. \qquad (55)$$

Since $DS_3(\alpha^0)$ does not depend on $h_\varphi(x)$, the second adjoint function [say $\lambda_3(x)$], which would have corresponded to $h_\varphi$, is identically zero, since the adjoint system for $\lambda_3(x)$ would have been a linear homogeneous equation with zero source and zero boundary conditions. In terms of the adjoint function $\theta_3(x)$, the expression of the differential $DS_3(\alpha^0)$ becomes:

$$\begin{aligned} DS_3(\alpha^0) &= \int_{-a}^{a} \theta_3(x)\left[(\delta D)\frac{d^2 \psi^0}{dx^2} - (\delta \Sigma_a)\psi^0(x) - (\delta \Sigma_d)\delta(x-b)\right]dx \\ &= S_{31}(\delta \Sigma_a) + S_{32}(\delta D) + S_{33}(\delta Q) + S_{34}(\delta \Sigma_d), \end{aligned} \qquad (56)$$

where

$$S_{31} \triangleq \frac{\partial^2 R}{\partial Q \partial \Sigma_a} = -\int_{-a}^{a} \theta_3(x)\psi^0(x)dx, \qquad (57)$$

$$S_{32} \triangleq \frac{\partial^2 R}{\partial Q \partial D} = \int_{-a}^{a} \theta_3(x) \frac{d^2 \psi^0}{dx^2} dx$$
$$= \left(\Sigma_a^0 / D^0\right) \int_{-a}^{a} \theta_3(x) \psi^0(x) dx + \left(\Sigma_d^0 / D^0\right) \int_{-a}^{a} \theta_3(x) \delta(x-b) dx, \tag{58}$$

$$S_{33} \triangleq \frac{\partial^2 R}{\partial Q \partial Q} = 0, \tag{59}$$

$$S_{34} \triangleq \frac{\partial^2 R}{\partial Q \partial \Sigma_d} = -\int_{-a}^{a} \theta_3(x) \delta(x-b) dx = -\theta_3(b). \tag{60}$$

Note that *the adjoint system comprising Eqs. (54) and (55) does not even need to be solved*, since a simple comparison of this system to the original diffusion Eq. (1) for the neutron flux $\varphi(x)$ shows that $\theta_3(x) = \varphi(x)/Q$. This fact can be readily verified by actually solving Eqs. (54) and (55) to obtain

$$\theta_3(x) = \frac{1}{\Sigma_a^0} \left[ \frac{\cosh(xk^0)}{\cosh(ak^0)} - 1 \right]. \tag{61}$$

Introducing the above expression for $\theta_3(x)$ into Eqs.(57) through (60), and performing the respective integrations leads to the following results for the second-order sensitivities $S_{3i}$:

$$S_{31}(\alpha^0) \triangleq \frac{\partial^2 R}{\partial Q \partial \Sigma_a} = -\frac{\Sigma_d^0}{\left(\Sigma_a^0\right)^2} A(k^0) + \frac{\Sigma_d^0}{2\Sigma_a^0 \sqrt{D^0 \Sigma_a^0}} B(k^0) \tag{62}$$

$$S_{32}(\alpha^0) \triangleq \frac{\partial^2 R}{\partial Q \partial D} = -\frac{\Sigma_d^0 k^0}{2D^0 \Sigma_a^0} B(k^0) \tag{63}$$

$$S_{34}(\alpha^0) \triangleq \frac{\partial^2 R}{\partial Q \partial \Sigma_d} = -\theta_3(b) = \frac{A(k^0)}{\Sigma_a^0} \tag{64}$$

Note that the computation of $S_{34}$ via Eq. (60) gives the same result as has already been obtained from the computation of $S_{43}$ via Eq. (50). This important consideration provides an *independent verification* that the adjoint functions $\theta_3(x)$ and, respectively, $\lambda_4(x)$, have indeed been computed correctly.

*2.2.3. Applying the SO-ASAP to Compute the Second-Order Response Sensitivities $S_{1i} \triangleq \partial^2 R/(\partial \Sigma_a \partial \alpha_i)$.*

The G-differential, $DS_1(\alpha^0)$, of the first-order sensitivity $S_1(\alpha)$ is computed by applying the definition given in Eq. (38) to Eq. (33), to obtain

$$DS_1(\alpha^0) = \frac{d}{d\varepsilon}\left\{\int_{-a}^{a}(\psi^0 + \varepsilon h_\psi)(\varphi^0 + \varepsilon h_\varphi)\,dx\right\}_{\varepsilon=0} = \int_{-a}^{a}\left(h_\varphi \psi^0 + h_\psi \varphi^0\right)dx \tag{65}$$

The function $h_\varphi$ is the solution of Eqs. (14) and (15), while $h_\psi$ is the solution of Eqs. (52) and (53). Note that although the entire contribution to $DS_1(\alpha^0)$ comes from the "indirect-effect" term [the "direct-effect" term contribution in Eq. (65) is identically zero], $DS_1(\alpha^0)$ does depend on both $h_\varphi$ and $h_\psi$. Consequently, as shown in the general theory presented in Part I, corresponding to $DS_1(\alpha^0)$ there will be *two non-zero adjoint systems and two adjoint functions*, $\lambda_1(x)$ and $\theta_1(x)$, which will be the solutions of the systems adjoint to those corresponding to the systems satisfied by $h_\varphi$ and, respectively, $h_\psi$. Thus, the adjoint system for $\lambda_1(x)$, corresponding to Eqs. (14) and (15) as well as to the response $DS_1(\alpha^0)$ is obtained as

$$D^0 \frac{d^2 \lambda_1}{dx^2} - \Sigma_a^0 \lambda_1(x) = \psi^0, \tag{66}$$

$$\lambda_1(\pm a) = 0, \tag{67}$$

while the adjoint system corresponding to Eqs. (52) and (53) as well as to the response $DS_1(\alpha^0)$ is obtained as

$$D^0 \frac{d^2\theta_1}{dx^2} - \Sigma_a^0 \theta_1(x) = \varphi^0, \tag{68}$$

$$\theta_1(\pm a) = 0. \tag{69}$$

In view of Eqs. (65) – (69), the G-differential $DS_1(\alpha^0)$ can be expressed in terms of the adjoint functions $\lambda_1(x)$ and $\theta_1(x)$ in the form

$$\begin{aligned}
DS_1(\alpha^0) &= \int_{-a}^{a} \lambda_1(x) \left[ -(\delta D) \frac{d^2\varphi^0}{dx^2} + (\delta \Sigma_a) \varphi^0(x) - (\delta Q) \right] dx \\
&+ \int_{-a}^{a} \theta_1(x) \left[ -(\delta D) \frac{d^2\psi^0}{dx^2} + (\delta \Sigma_a) \psi^0(x) + (\delta \Sigma_d) \delta(x-b) \right] dx \\
&= S_{11}(\delta \Sigma_a) + S_{12}(\delta D) + S_{13}(\delta Q) + S_{14}(\delta \Sigma_d),
\end{aligned} \tag{70}$$

where

$$S_{11}(\alpha^0) \triangleq \frac{\partial^2 R}{\partial \Sigma_a \partial \Sigma_a} = \int_{-a}^{a} \left[ \lambda_1(x) \varphi^0(x) + \theta_1(x) \psi^0(x) \right] dx, \tag{71}$$

$$\begin{aligned}
S_{12}(\alpha^0) \triangleq \frac{\partial^2 R}{\partial D \partial \Sigma_a} &= -\int_{-a}^{a} \left[ \lambda_1(x) \frac{d^2\varphi^0}{dx^2} + \theta_1(x) \frac{d^2\psi^0}{dx^2} \right] dx \\
&= -(\Sigma_a^0/D) \int_{-a}^{a} \left[ \lambda_1(x) \varphi^0(x) + \theta_1(x) \psi^0(x) \right] dx + (Q^0/D) \int_{-a}^{a} \lambda_1(x) dx \\
&\quad - (\Sigma_d^0/D) \int_{-a}^{a} \theta_1(x) \delta(x-b) dx,
\end{aligned} \tag{72}$$

$$S_{13}(\alpha^0) \triangleq \frac{\partial^2 R}{\partial Q \partial \Sigma_a} = -\int_{-a}^{a} \lambda_1(x) dx, \tag{73}$$

$$S_{14}(\alpha^0) \triangleq \frac{\partial^2 R}{\partial \Sigma_d \partial \Sigma_a} = \int_{-a}^{a} \theta_1(x)\delta(x-b)dx = \theta_1(b). \tag{74}$$

The solution $\lambda_1(x)$ of adjoint system shown in Eqs. (66) and (67) is

$$\lambda_1(x) = \lambda_{1p}(x) - \sinh(x)\frac{\lambda_{1p}(a) - \lambda_{1p}(-a)}{2\sinh(ak^0)} - \cosh(x)\frac{\lambda_{1p}(a) + \lambda_{1p}(-a)}{2\cosh(ak^0)}, \tag{75}$$

with

$$\begin{aligned}\lambda_{1p}(x) =& \frac{\Sigma_d^0}{D^0 \Sigma_a^0} \frac{\sinh(bk^0 - ak^0)}{\cosh(ak^0)} \\ &\times \left\{\frac{x}{2}\cosh(xk^0 + ak^0) + \frac{1}{4k^0}\cosh(2xk^0 + ak^0)\left[\sinh(xk^0) - \cosh(xk^0)\right]\right\} \\ &- \frac{\Sigma_d^0}{D^0 \Sigma_a^0} H(x-b) \\ &\times \left\{\frac{x}{2}\cosh(xk^0 - bk^0) + \frac{1}{4k^0}\cosh(2xk^0 - bk^0)\left[\sinh(xk^0) - \cosh(xk^0)\right]\right\}\end{aligned} \tag{76}$$

Similarly, solving the adjoint system consisting of Eqs.(68) and (69) yields

$$\begin{aligned}\theta_1(x) =& \frac{Q^0}{2\Sigma_a^0\sqrt{D^0\Sigma_a^0}} \frac{a\sinh(ak^0)\cosh(xk^0) - x\sinh(xk^0)\cosh(ak^0)}{\cosh^2(ak^0)} \\ &+ \frac{Q^0}{(\Sigma_a^0)^2}\left[\frac{\cosh(xk^0)}{\cosh(ak^0)} - 1\right].\end{aligned} \tag{77}$$

The second-order sensitivity $S_{14}(\alpha^0)$ can now be computed from Eqs. (77) and (74) to obtain the same expression as was obtained in Eq. (48) for $S_{41}(\alpha^0)$, namely:

$$S_{14}(\alpha^0) = \theta_1(b) = S_{41}(\alpha^0) = Q^0\left[-(\Sigma_a^0)^{-2} A(k^0) + \frac{1}{2\Sigma_a^0\sqrt{D^0\Sigma_a^0}} B(k^0)\right]. \tag{48}$$

The above result provides an independent verification that the adjoint functions $\theta_1(x)$ and $\lambda_4(x)$ have been computed correctly. Next, inserting the expressions of $\lambda_1(x)$ and $\theta_1(x)$ into Eqs. (71) through (73), and performing the respective integrations, yields the following expressions:

$$S_{11}(\alpha^0) \triangleq \frac{\partial^2 R}{\partial \Sigma_a^2} = \frac{Q^0 \Sigma_d^0}{\left(\Sigma_a^0\right)^3}\left[2A(k^0) - \frac{5}{4}k^0 B(k^0) + \frac{1}{4}(k^0)^2 C(k^0)\right] \tag{78}$$

$$S_{21}(\alpha^0) \triangleq \frac{\partial^2 R}{\partial D \partial \Sigma_a} = \frac{Q^0 \Sigma_d^0}{4D^0 \left(\Sigma_a^0\right)^2} k^0 \left[B(k^0) - k^0 C(k^0)\right] \tag{79}$$

$$S_{31}(\alpha^0) \triangleq \frac{\partial^2 R}{\partial Q \partial \Sigma_a} = S_{13}(\alpha^0) = \frac{\Sigma_d^0}{\Sigma_a^0}\left[-\frac{A(k^0)}{\Sigma_a^0} + \frac{B(k^0)}{2\sqrt{D^0 \Sigma_a^0}}\right] \tag{62}$$

where

$$C(k^0) \triangleq \left.\frac{dB}{dk}\right|_{k^0} = \left.\frac{d^2 A}{dk^2}\right|_{k^0} = 2a^2 \cosh(bk^0)\left[\cosh(ak^0)\right]^{-3}$$
$$+ 2ab\sinh(ak^0)\sinh(bk^0)\left[\cosh(ak^0)\right]^{-2} - (a^2+b^2)\cosh(bk^0)\left[\cosh(ak^0)\right]^{-1} \tag{80}$$

Of course, the expression of $S_{31}(\alpha^0) = S_{13}(\alpha^0)$ has already been computed in Eq. (62), which is the reason why it was accordingly labeled above, so its computation from Eq. (73) serves only as an additional independent verification that the adjoint function $\lambda_1(x)$ has been correctly computed. This verification is in the same spirit as the verification that was performed on the adjoint functions $\theta_1(x)$ when computing $S_{14}(\alpha^0)$ from Eq. (74) and then ensuring that the resulting expression was identical to that already obtained in Eq. (48) for $S_{41}(\alpha^0)$.

On the other hand, the expressions of $S_{11}(\alpha^0)$ and $S_{21}(\alpha^0)$ obtained in Eqs. (78) and (79), respectively, are in addition to the expressions already obtained for the second-order sensitivities in Sections 2.2.1 and 2.2.2.

*2.2.4. Applying the SO-ASAP to Compute the Second-Order Response Sensitivities* $S_{2i} \triangleq \partial^2 R/(\partial D \partial \alpha_i)$.

The G-differential, $DS_2(\alpha^0)$, of the first-order sensitivity $S_2(\alpha)$ is computed by applying the the definition given in Eq. (38) to Eq. (34), to obtain

$$DS_2(\alpha^0) = \frac{d}{d\varepsilon} \left\{ -\frac{\left[\Sigma_a^0 + \varepsilon(\delta\Sigma_a)\right]}{D^0 + \varepsilon(\delta D)} \int_{-a}^{a} (\psi^0 + \varepsilon h_\psi)(\varphi^0 + \varepsilon h_\varphi) dx + \left[Q^0 + \varepsilon(\delta Q)\right] \int_{-a}^{a} (\psi^0 + \varepsilon h_\psi) dx \right\}_{\varepsilon=0}$$
$$= \{DS_2(\alpha^0)\}_{direct} + \{DS_2(\alpha^0)\}_{indirect},$$
(81)

where

$$\{DS_2(\alpha^0)\}_{direct} \triangleq \frac{(\delta D)\Sigma_a^0 - (\delta\Sigma_a)D^0}{(D^0)^2} \int_{-a}^{a} \psi^0(x)\varphi^0(x) dx$$
$$+ \frac{D^0(\delta Q) - (\delta D)Q^0}{(D^0)^2} \int_{-a}^{a} \psi^0(x) dx,$$
(82)

and

$$\{DS_2(\alpha^0)\}_{indirect} \triangleq \int_{-a}^{a} h_\varphi(x) \left[-\frac{\Sigma_a^0}{D^0}\psi^0(x)\right] dx + \int_{-a}^{a} h_\psi(x) \left[\frac{Q^0}{D^0} - \frac{\Sigma_a^0}{D^0}\varphi^0(x)\right] dx.$$
(83)

In Eqs. (81) through (83), the function $h_\varphi$ is the solution of Eqs. (14) and (15), while $h_\psi$ is the solution of Eqs. (52) and (53), just as was the case in the previous sections. Note that, in contradistinction to the G-differentials $DS_{i \neq 2}(\alpha^0)$, the G-differential $DS_2(\alpha^0)$ comprises not only an "indirect-effect" term, but also a (non-zero) "direct-effect" term. The "direct-effect" term can be evaluated immediately since all of the quantities in Eq. (82) are already available. On the other hand, the "indirect-effect" term $\{DS_2(\alpha^0)\}_{indirect}$ depends on both $h_\varphi$ and $h_\psi$, so that *two non-zero adjoint functions*, labeled here as $\lambda_2(x)$ and $\theta_2(x)$, are needed in order to evaluate it. These adjoint functions will be solutions corresponding to the systems adjoint to those satisfied by $h_\varphi$ and $h_\psi$, respectively, but with sources derived from the representation of $\{DS_2(\alpha^0)\}_{indirect}$ in Eq. (83). Applying by now the familiar *SO-ASAP* yields the *second adjoint system for* $\lambda_2(x)$ [which corresponds to Eqs. (14) and (15), as well as to the response in Eq. (83)], of the form

$$D^0 \frac{d^2 \lambda_2}{dx^2} - \Sigma_a^0 \lambda_2(x) = -\frac{\Sigma_a^0}{D^0} \psi^0, \tag{84}$$

$$\lambda_2(\pm a) = 0, \tag{85}$$

together the *second adjoint system for* $\theta_2(x)$ [which corresponds to Eqs. (52) and (53), as well as to the response in Eq. (83)], of the form

$$D^0 \frac{d^2 \theta_2}{dx^2} - \Sigma_a^0 \theta_2(x) = \frac{Q^0}{D^0} - \frac{\Sigma_a^0}{D^0} \varphi^0, \tag{86}$$

$$\theta_2(\pm a) = 0. \tag{87}$$

The G-differential $\{DS_2(\alpha^0)\}_{indirect}$ can now be expressed in terms of the adjoint functions $\lambda_2(x)$ and $\theta_2(x)$ as

$$\begin{aligned}\{DS_2(\alpha^0)\}_{indirect} &= \int_{-a}^{a} \lambda_2(x)\left[-(\delta D)\frac{d^2\varphi^0}{dx^2} + (\delta\Sigma_a)\varphi^0(x) - (\delta Q)\right]dx \\ &+ \int_{-a}^{a} \theta_2(x)\left[-(\delta D)\frac{d^2\psi^0}{dx^2} + (\delta\Sigma_a)\psi^0(x) + (\delta\Sigma_d)\delta(x-b)\right]dx.\end{aligned} \tag{88}$$

Combinig the above result with the "direct-effect" term, $\{DS_2(\alpha^0)\}_{direct}$, defined in Eq. (82), yields:

$$DS_2(\alpha^0) = S_{21}(\delta\Sigma_a) + S_{22}(\delta D) + S_{23}(\delta Q) + S_{24}(\delta\Sigma_d), \tag{89}$$

where

$$S_{21}(\alpha^0) \triangleq \frac{\partial^2 R}{\partial D \partial \Sigma_a} = (-1/D^0) \int_{-a}^{a} \varphi^0(x)\psi^0(x)dx + \int_{-a}^{a} \lambda_2(x)\varphi^0(x)dx$$
$$+ \int_{-a}^{a} \theta_2(x)\psi^0(x)dx, \tag{90}$$

$$S_{22}(\alpha^0) \triangleq \frac{\partial^2 R}{\partial D^2} = \frac{\Sigma_a^0}{(D^0)^2} \int_{-a}^{a} \varphi^0(x)\psi^0(x)dx - \frac{Q^0}{(D^0)^2} \int_{-a}^{a} \varphi^0(x)\psi^0(x)dx$$
$$\int_{-a}^{a} \left[ \lambda_2(x)\frac{d^2\varphi^0}{dx^2} + \theta_2(x)\frac{d^2\psi^0}{dx^2} \right]dx, \tag{91}$$

$$S_{23}(\alpha^0) \triangleq \frac{\partial^2 R}{\partial D \partial Q} = (1/D^0) \int_{-a}^{a} \psi^0(x)dx - \int_{-a}^{a} \lambda_2(x)dx, \tag{92}$$

$$S_{24}(\alpha^0) \triangleq \frac{\partial^2 R}{\partial D \partial \Sigma_d} = \int_{-a}^{a} \theta_2(x)\delta(x-b)dx = \theta_2(b). \tag{93}$$

To compute the last integral on the right-side of Eq. (91), we use Eqs. (1) and (27) to obtain

$$\lambda_2(x)\frac{d^2\varphi^0}{dx^2} + \theta_2(x)\frac{d^2\psi^0}{dx^2} = \lambda_2(x)\left[\frac{\Sigma_a^0}{D^0}\varphi^0(x) - \frac{Q^0}{D^0}\right] + \theta_2(x)\left[\frac{\Sigma_a^0}{D^0}\psi^0(x) + \frac{\Sigma_d^0}{D^0}\delta(x-b)\right],$$

and subsequently replace the right-side of the above expression in the last integral on the right-side of Eq. (91).

Comparing Eqs. (84) and (85) with Eqs. (66) and (67) readily reveals that the sources of these two linear systems are proportional to each other by the factor $(-\Sigma_a^0/D^0)$. Hence, the solutions of these two systems will be proportional to each other by the same factor. Consequently, *Eqs. (84) and (85) need not even be solved* since the solution $\lambda_2(x)$ can be immediately written down in terms of the $\lambda_1(x)$ as

$$\lambda_2(x) = \left(-\Sigma_a^0/D^0\right)\lambda_1(x) \tag{94}$$

with the expression of $\lambda_1(x)$ provided in Eq. (75). On the other hand, the solution $\theta_2(x)$ of Eqs. (86) and (87) is obtained by conventional methods as

$$\theta_2(x) = \frac{Q^0}{2k^0\left(D^0\right)^2} \frac{x\sinh\left(xk^0\right)\cosh\left(ak^0\right) - a\sinh\left(ak^0\right)\cosh\left(xk^0\right)}{\cosh^2\left(ak^0\right)}. \tag{95}$$

Since the expression of the second-order sensitivity $S_{42}(\alpha^0)$ is already available from Eq. (49), it follows that Eq. (93) provides an *exact verification* of the adjoint function $\theta_2(x)$. This is indeed the case, as demonstrated by the fact that

$$S_{24}(\alpha^0) = S_{42}(\alpha^0) = \theta_2(b) \tag{96}$$

The expression of $S_{21}(\alpha^0)$ has already been obtained in Eq. (79), and the expression of $S_{32}(\alpha^0)$ has already been obtained in Eq. (63). Therefore, Eqs. (90) through (92) can serve as multiple verifications of the correctness of the various computations. This is indeed the case: inserting the expressions given in Eqs. (94) and (95) [for $\lambda_2(x)$ and $\theta_2(x)$, respectively] together with the Eqs. (3) and (30) [i.e., the expressions of $\varphi^0(x)$ and $\psi^0(x)$] into Eqs. (90) through (92), and performing the respective intergrations, leads to the following results:

$$S_{21}(\alpha^0) = S_{12}(\alpha^0) = \frac{Q^0 \Sigma_d^0}{4D^0\left(\Sigma_a^0\right)^2} k^0 \left[B\left(k^0\right) - k^0 C\left(k^0\right)\right], \tag{97}$$

and

$$S_{23}(\alpha^0) = S_{32}(\alpha^0) = -\frac{\Sigma_d^0 k^0}{2D^0 \Sigma_a^0} B\left(k^0\right). \tag{98}$$

The only new result which is actually obtained when computing $DS_2(\alpha^0)$, after all of the other differentials $DS_{i \neq 2}(\alpha^0)$ have already been computed, is the expression for the second-order sensitivity $S_{22}(\alpha^0)$. Thus, performing the integrations in Eq. (91) yields

$$S_{22}(\alpha^0) \triangleq \frac{\partial^2 R}{\partial D^2} = \frac{Q^0 \Sigma_d^0}{4(D^0)^2} \left[ 3 \frac{k^0}{\Sigma_a^0} B(k^0) + \frac{1}{D^0} C(k^0) \right]. \tag{99}$$

Having now finished the computation of all of the second-order response sensitivities, it is instructive to count the number of adjoint systems that needed to be solved, since these computations would be the equivalent of the "large-scale" computations performed if the *SO-ASAP* were used in practice. For one functional-type response, the count is as follows:

(i) *one adjoint computation* to determine the function $\psi^0(x)$, which suffices to compute, using just quadratures, all of the first-order sensitivities $S_i \equiv \partial R / \partial \alpha_i$, $1 = 1, 2, 3, 4$; and for subsequently determining (through a trivial division) all of the second-order sensitivities $S_{4i} \equiv \partial^2 R / \partial Q \partial \alpha_i$, $1 = 1, 2, 3, 4$,

(ii) *two adjoint computations* for determining $S_{1i} \equiv \partial^2 R / \partial \Sigma_a \partial \alpha_i$, $1 = 1, 2$;

(iii)   *one adjoint computation* for determining $S_{22} \equiv \partial^2 R / \partial D^2$.

To summarize: for each response, *four (4) "large-scale" adjoint computations sufficed for the complete and exact computations of all (4) first- and (10) distinct second-order derivatives, including the various verifications of the adjoint functions and derivative-symmetry properties.* Also, for all of the adjoint computations, *only the sources on the right-side of Eq. (1) [i.e., the initial diffusion equation for the neutron flux $\varphi^0(x)$] needed modifications.* The actual solver for the diffusion equation remained unchanged. By comparison, forward methods, e.g., the *SO-FSAP,* require 14 "large-scale" forward computations (i.e., solutions of the diffusion equation) for computing all of the first- and (distinct) second-order sensitivities.

It is instructive to compute the second-order relative sensitivities using the same data (and for the same responses) as used in Section 2.2 for the computations of the first-order response sensitivities. The corresponding numerical results are presented in Table 3 (second-order absolute sensitivities, with the corresponding units ommitted) and Table 4 (second-order relative sensitivities).

Table 3. Second-order (absolute values, sans units) sensitivities for six detector responses

| 2$^{nd}$ Order Absolute Sensitivities | $R_1$ (10 cm) $R_4$ (-10 cm) | $R_2$ (40 cm) $R_5$ (-40 cm) | $R_3$ (49.5 cm) $R_6$ (-49.5 cm) |
|---|---|---|---|
| $S_{11}$ | 1.95x10$^{13}$ | 1.67x10$^{13}$ | 1.28x10$^{12}$ |
| $S_{12}$ | 5.08x10$^{7}$ | 1.42x10$^{11}$ | 5.18x10$^{10}$ |
| $S_{13}$ | -1.92x10$^{4}$ | -1.76x10$^{4}$ | -1.67x10$^{3}$ |
| $S_{14}$ | -2.58x10$^{10}$ | -2.36x10$^{10}$ | -2.25x10$^{9}$ |
| $S_{22}$ | -4.59x10$^{6}$ | -1.97x10$^{9}$ | 1.53x10$^{10}$ |
| $S_{23}$ | -1.33x10$^{-2}$ | -1.24x10$^{2}$ | -1.74x10$^{2}$ |
| $S_{24}$ | -1.79x10$^{4}$ | -1.67x10$^{8}$ | -2.33x10$^{8}$ |
| $S_{33}$ | 0 | 0 | 0 |
| $S_{34}$ | 5.08x10$^{1}$ | 4.92x10$^{1}$ | 8.16 |
| $S_{44}$ | 0 | 0 | 0 |

Table 4. Second-order (relative) sensitivities for six detector responses

| 2$^{nd}$ Order Relative Sensitivities | $R_1$ (10 cm) $R_4$ (-10 cm) | $R_2$ (40 cm) $R_5$ (-40 cm) | $R_3$ (49.5 cm) $R_6$ (-49.5 cm) |
|---|---|---|---|
| $S_{11}$_rel | 2.00 | 1.77 | 0.82 |
| $S_{12}$_rel | 4.24x10$^{-5}$ | 0.12 | 0.27 |
| $S_{13}$_rel | -1.00 | -0.95 | -0.54 |
| $S_{14}$_rel | -1.00 | -0.95 | -0.54 |
| $S_{22}$_rel | -3.11x10$^{-5}$ | -0.014 | 0.65 |
| $S_{23}$_rel | -5.64x10$^{-6}$ | -0.054 | -0.46 |
| $S_{24}$_rel | -5.64x10$^{-6}$ | -0.054 | -0.46 |
| $S_{33}$_rel | 0 | 0 | 0 |
| $S_{34}$_rel | 1.00 | 1.00 | 1.00 |
| $S_{44}$_rel | 0 | 0 | 0 |

The values of the absolute second-order sensitivities presented in Table 3 will be used in the following Section to illustrate their essential role for quantifying non-Gaussian features (e.g., asymmetries) of the various response distributions. To quantify assymetries in distribution, at the very least the third-order ("skewness") response correlations need to be computed, which require the exact computation of (at least) the first- and second-order response sensitivities to model parameters. Also, we note from Table 4 that, "conventional folklore" (which is to argue that "second-order sensitivities are sufficiently small to be neglected") potentially ignores significant information about the system under study.in reactor physics, The results in Table 4 clearly indicate that second-order relative sensitivities can be just as large as (or even significantly larger than) first-order sensitivities.

## *2.3. Illustration of the Essential Role Played by the Second-Order Response Sensitivities for Quantifying Non-Gaussian Features of the Response Uncertainty Distribution*

In general, the model parameters are experimentally derived quantities and are therefore subject to uncertainties. Specifically, consider that the model comprises $N_\alpha$ uncertain parameters $\alpha_i$, which constitute the components of the (column) vector $\boldsymbol{\alpha}$ of *model parameters*, defined as $\boldsymbol{\alpha} = (\alpha_1, ..., \alpha_{N_\alpha})$. The usual information available in practice comprises the mean values of the model parameters together with uncertainties (standard deviations and, occasionally, correlations) computed about the respective mean values. The components of vector $\boldsymbol{\alpha}^0 \equiv (\alpha_1^0, ..., \alpha_{N_\alpha}^0)$ of mean values of the model parameters are denoted as $\alpha_i^0$ and defined as

$$\alpha_i^0 \equiv \langle \alpha_i \rangle, \quad \langle f \rangle \equiv \int f(\boldsymbol{\alpha}) p(\boldsymbol{\alpha}) d\boldsymbol{\alpha}, . \tag{100}$$

where the angular brackets denotes integration of a generic function $f(\boldsymbol{\alpha})$ over the *unknown* joint probability distribution, $p(\boldsymbol{\alpha})$, of the parameters $\boldsymbol{\alpha}$. The parameter distribution's second-order central moments, $\mu_2^{ij}(\boldsymbol{\alpha})$, are defined as

$$\mu_2^{ij}(\boldsymbol{\alpha}) \equiv \left\langle (\alpha_i - \alpha_i^0)(\alpha_j - \alpha_j^0) \right\rangle \equiv \rho_{ij}\sigma_i\sigma_j; \quad i, j = 1, ..., N_\alpha. \tag{101}$$

The central moments $\mu_2^{ii}(\boldsymbol{\alpha}) \equiv \text{var}(\alpha_i)$ are called the *variance* of $\alpha_i$, while the central moments $\mu_2^{ij}(\boldsymbol{\alpha}) \equiv \text{cov}(\alpha_i, \alpha_j); i \neq j$, are called the *covariances* of $\alpha_i$ and $\alpha_j$. The *standard deviation* of $\alpha_i$ is defined as $\sigma_i \equiv \sqrt{\mu_2^{ii}(\boldsymbol{\alpha})}$. For a (univariate) distribution of a variate $\alpha_i$, the expected (or mean) value $\langle \alpha_i \rangle$ is a measure of the *location* of the respective distribution, while the standard deviation $\sigma_i \equiv \sqrt{\mu_2^{ii}(\boldsymbol{\alpha})}$ provides a measure for the *dispersion* of the respective distribution around the mean value. In the same sense, the *mean value can be interpreted as the distribution's center of gravity*, while the variance can be interpreted as the distribution's moment of inertia (whoich in mechanics linearly relates the applied torque to the induced acceleration).

When the model under consideration is used to compute $N_r$ *responses* (or results), denoted in vector form as $\mathbf{r} = (r_1, ..., r_{N_r})$, each of these responses will be implicit functions of the model's parameters, i.e., $\mathbf{r} = \mathbf{r}(\boldsymbol{\alpha})$. It follows that $\mathbf{r} = \mathbf{r}(\boldsymbol{\alpha})$ will be a vector-valued variate which obeys a (generally intractable) multivariate distribution in $\boldsymbol{\alpha}$. For large-scale systems, the probability distribution $p(\boldsymbol{\alpha})$ is not known in practice and, even if it were known, the induced distribution in $\mathbf{r} = \mathbf{r}(\boldsymbol{\alpha})$ would still be intractable, since $p(\boldsymbol{\alpha})$ could not be propagated exactly through the large-scale models used in for simulating realistic physical systems.

The time-honored deterministic method for computing uncertainties in a response $\mathbf{r}(\boldsymbol{\alpha})$ arising from uncertainties in the parameters $\boldsymbol{\alpha}$ relies on expanding formally the response $\mathbf{r}(\boldsymbol{\alpha})$ in a Taylor series around $\boldsymbol{\alpha}^0$, constructing appropriate products of such Taylor series, and intergrating formally the various products over the unknown parameter distribution function $p(\boldsymbol{\alpha})$, to obtain response correlations. This method for constructing response correlations stemming from parameter correlations is known as the "propagation of errors" or "propagation of moments" method. Using this method, reference [5] provides expressions for response correlations up to fourth-order, using up to fourth-order parameter correlations and response sensitivities.

For illustrating the effects of second-order response sensitivities for the paradigm neutron diffusion problem considered in this work, it suffices to take from Ref. [5] response correlations up to third-order, for the very simple case when: (i) the paranmeters are uncorrelated and normally distributed; and (ii) only the first- and second-order response sensitivities are available. For these particular conditions, the response correlations derived in [5] reduce to the following expressions for the first three response moments:

(i) The expected value of a response $r_k$, denoted here as $\left[E(r_k)\right]^{UG}$, which arises due to uncertainties in uncorrelated normally-distributed model parameters (the superscript *UG* indicates "uncorrelated Gaussian" parameters), is given by the expression

$$\left[E(r_k)\right]^{UG} = r_k(\boldsymbol{\alpha}^0) + \frac{1}{2}\sum_{i=1}^{N_\alpha} \frac{\partial^2 r_k}{\partial \alpha_i^2} \sigma_i^2 , \qquad (102)$$

where $r_k(\boldsymbol{\alpha}^0)$ denotes the computed nominal value of the response;

(ii) The covariance, $cov(r_k, r_\ell)$, between two responses $r_k$ and $r_\ell$ arising from normally-distributed uncorrelated parameters is given by

$$\left[cov(r_k, r_\ell)\right]^{UG} = \sum_{i=1}^{N_\alpha} \left(\frac{\partial r_k}{\partial \alpha_i}\frac{\partial r_\ell}{\partial \alpha_i}\right)\sigma_i^2 + \frac{1}{2}\sum_{i=1}^{N_\alpha}\left(\frac{\partial^2 r_k}{\partial \alpha_i^2}\right)\left(\frac{\partial^2 r_\ell}{\partial \alpha_i^2}\right)\sigma_i^4 \qquad (103)$$

The variance, $var(r_k)$, of a response $r_k$ is obtained by setting $r_k \equiv r_l$ in the above expression to obtain

$$\left[var(r_k)\right]^{UG} = \sum_{i=1}^{N_\alpha}\left(\frac{\partial r_k}{\partial \alpha_i}\right)^2 \sigma_i^2 + \frac{1}{2}\sum_{i=1}^{N_\alpha}\left(\frac{\partial^2 r_k}{\partial \alpha_i^2}\right)^2 \sigma_i^4 \; ; \qquad (104)$$

(iii) The third-order response correlation, $\mu_3(r_k, r_l, r_m)$, among three responses ($r_k$, $r_\ell$ and $r_m$) has the following expression:

$$\left[\mu_3(r_k, r_l, r_m)\right]^{UG} = \sum_{i=1}^{N_\alpha}\left(\frac{\partial^2 r_k}{\partial \alpha_i^2}\frac{\partial r_l}{\partial \alpha_i}\frac{\partial r_m}{\partial \alpha_i} + \frac{\partial r_k}{\partial \alpha_i}\frac{\partial^2 r_l}{\partial \alpha_i^2}\frac{\partial r_m}{\partial \alpha_i} + \frac{\partial r_k}{\partial \alpha_i}\frac{\partial r_l}{\partial \alpha_i}\frac{\partial^2 r_m}{\partial \alpha_i^2}\right)\sigma_i^4. \qquad (105)$$

In particular, the third-order central moment, $\mu_3(r_k)$, of the response $r_k$ is determined by setting $k = l = m$ in Eq. (105) to obtain

$$\left[\mu_3(r_k)\right]^{UG} = 3\sum_{i=1}^{N_\alpha}\left(\frac{\partial r_k}{\partial \alpha_i}\right)^2 \frac{\partial^2 r_k}{\partial \alpha_i^2}\sigma_i^4 \qquad (106)$$

The *skewness*, $\gamma_1(r_k)$, of a response $r_k$ can be computed using the customary definition

$$\gamma_1(r_k) \equiv \mu_3(r_k)\big/\left[var(r_k)\right]^{3/2}. \qquad (107)$$

The *skewness* of a distribution quantifies the departure of the subject distribution from symmetry. Symmetric univariate distributions (e.g., the Gaussian) are characterized by $\gamma_1(r_k) = 0$. A distribution with a long right tail would have a positive skewness while a distribution with a long left tail would have a negative skewness. In other words, if $\gamma_1(r_k) < 0$, then the respective distribution is skewed towards the left of the mean $[E(r_k)]^{UG}$, favoring lower values of $r_k$ relative to $[E(r_k)]^{UG}$. On the other hand, if $\gamma_1(r_k) > 0$, then the respective distribution is skewed towards the right of the mean $[E(r_k)]^{UG}$, favoring higher values of $r_k$ relative to $[E(r_k)]^{UG}$.

As indicated by the expressions in Eqs. (102) - (107), the second-order sensitivities have the following impacts on the response moments:
   (a) They cause the "expected value of the response", $[E(r_k)]^{UG}$, to differ from the "computed nominal value of the response", $r_k(\boldsymbol{\alpha}^0)$;
   (b) They contribute to the response variances and covariances; however, since the contributions involving the second-order sensitivities are multiplied by the fourth power of the parameters' standard deviations, the total of these contributions is expected to be relatively smaller than the contributions stemming from the first-order response sensitivities;
   (c) They *contribute decisively to causing assymetries in the response distribution*. As Eq. (105) indicates, neglecting the second-order sensitivities would nullify the third-order response correlations and hence would nullify the *skewness*, $\gamma_1(r_k)$, of a response $r_k$, cf. Eq. (107). Consequently, any events occurring in a response's long and/or short tails, which are characteristic of rare but decisive events (e.g., major accidents, catastrophies), would likely be missed.

To illustrate the above points numerically for our paradigm neutron diffusion problem, it is convenient to recall here that: $a = 50\,cm$, $S^0 = 10^7\,neutrons \cdot cm^{-3} \cdot s^{-1}$, $\Sigma_a^0 = 0.0197\,cm^{-1}$, $D^0 = 0,16\,cm$, $\Sigma_d^0 = 7.438\,cm^{-1}$. Recall also that the computed responses had the following values in units of $[neutrons \cdot cm^{-3} \cdot s^{-1}]$ : $R_1(10\,cm) = R_4(-10\,cm) = 3.77 \times 10^9$, $R_2(40\,cm) = R_5(-40\,cm) = 3.66 \times 10^9$, and $R_3(49.5\,cm) = R_6(-49.5\,cm) = 6.076 \times 10^8$.

We will now investigate the effects of parameter uncertainties by using Eqs. (102) – (107) to quantify the resulting uncertainties induced in the respected responses. Several combinations of parameters' relative standard deviations, denoted as $\delta\Sigma_A(\%)$, $\delta D(\%)$, $\delta Q(\%)$, and $\delta\Sigma_D(\%)$ will be considered for this purpose; the various cases and corresponding relative standard deviations are listed in Table 5.

Table 5. Model parameters' relative standard deviations

| Case | $\delta\Sigma_a(\%)$ | $\delta D(\%)$ | $\delta Q(\%)$ | $\delta\Sigma_d(\%)$ |
|---|---|---|---|---|
| 1 | 0 | 0 | 15 | 0 |
| 2 | 0 | 0 | 0 | 15 |
| 3 | 15 | 0 | 0 | 0 |
| 4 | 0 | 15 | 0 | 0 |
| 5 | 10 | 10 | 10 | 10 |

The expression given in Eq. (104) has been used to compute the standard deviations induced in responses by the parameter uncertainties listed in Table 5. The results of these computations are shown in Table 6. Examination of cases 1 and 2 indicates that uncertainties arising solely in $Q$ and $\Sigma_d$, respectively, propagate "one-to-one", inducing corresponding uncertainties in all of the responses; this fact is expected in view of their first-order relative sensitivities of unity, for all responses (as shown in Table 2). The uncertainties in $\Sigma_a$ propagate almost one-to-one to produce the response uncertainties in the interior of the slab, but cause diminishingly smaller uncertainties in the detector response towards the slab's boundaries; again, this behavior is expected in view of the values for the corresponding first-order sensitivities shown in Table 2. The impact of the uncertainties in $D$ on the various responses is also dictated by the behavior of the corresponding first-order sensitivities shown in Table 2: the uncertainties in $D$ have a negligible impact on the responses in the interior of the slab, but their impact increases as the detector responses approach the medium's boundaries. Finally, when all of the parameters are uncertain, their combined contribution to the respective response uncertainty is larger than any one parameter's uncertainty, but is smaller than the "the square-root of the sums of squares", because the response sensitivities to $\Sigma_a$ and $D$ are not unity. We also mention that the results (which are not shown, for brevity) of using Eq. (103) to compute the response correlations (value ranges 0.998 to 1.000) indicate responses are fully (or almost fully) correlated. In all of these computations, *the second-order sensitivities had a negligible impact on the responses' standard deviations and correlations.*

Table 6. Response relative standard deviations

| Case | $\sigma_{rel}(R_1) = \sigma_{rel}(R_4)$ | $\sigma_{rel}(R_2) = \sigma_{rel}(R_5)$ | $\sigma_{rel}(R_3) = \sigma_{rel}(R_6)$ |
|---|---|---|---|
| 1 | 0.150 | 0.150 | 0.150 |
| 2 | 0.150 | 0.150 | 0.150 |
| 3 | 0.150 | 0.145 | 0.082 |
| 4 | $9.8 \times 10^{-7}$ | 0.008 | 0.069 |
| 5 | 0.174 | 0.171 | 0.158 |

As indicated by the results presented in Table 7 for the skewness of the various responses, the second-order sensitivities have a decisive impact on the assymetries of the induced response distributions. Since the second-order sensitivities of the responses to both $Q$ and $\Sigma_d$ are zero, cf. Table 3, it follows that uncertainties arising solely in these quantities would not affect the symmetry of the resulting response distribution in our illustrative paradigm problem. This expectation is indeed confirmed by the results presented under "Case 1" and "Case 2" in Table 7. On the other hand, "Case 3" in Table 7 indicates that uncertainties stemming solely from the absoption cross section $\Sigma_a$ cause the resulting response distributions to be significantly skewed to the right of their corresponding expected values, throughout the medium; this asymmetry diminishes somewhat towards the medium's boundaries. Case "4" indicates a somewhat counter-intuitive behavior of the response distributions, in that uncertainties arising solely from the diffusion coefficient $D$ cause the response distribution to undergo a complex behavior, as follows:
   (i) the response distribution is skewed significantly to the left of the response's expected value in the medium's interior;
   (ii) the above-mentioned asymmetry diminishes towards the medium's boundary;

(iii) the response distribution become symmetric close to the boundary, and then dramatically "shifts directions" by becoming markedly skewed to the right in the "boundary-layer" close to the medium's physical boundary.

Finally, "Case 5" in Table 7 indicates the "compensatory effects" which are induced by similar uncertainties in $\Sigma_a$ and D, respectively. Similar standard deviations in $\Sigma_a$ and D induce opposite assymetries in the response distributions but collectively they tend to dimish the opposing effects. Overall, the distribution remains somewhat skewed to the right of the responses' respective mean values, throughout the medium.

Of course, none of the effects implied by the results shown in Table 7 would have been observed in the absence of the second-order response sensitivities.

Table 7. Response skewness

| Case | $\gamma_1(R_1)=\gamma_1(R_5)$ | $\gamma_1(R_2)=\gamma_1(R_6)$ | $\gamma_1(R_3)=\gamma_1(R_7)$ |
|---|---|---|---|
| 1 | 0 | 0 | 0 |
| 2 | 0 | 0 | 0 |
| 3 | 0.8425 | 0.7945 | 0.6519 |
| 4 | -1.5965 | -0.1143 | 0.6147 |
| 5 | 0.1143 | 0.0955 | 0.0283 |

## 3. SUMMARY AND CONCLUSIONS

This work has presented an illustrative application of the *second-order adjoint sensitivity analysis procedure (SO-ASAP)* [1] to a paradigm neutron diffusion problem which is sufficiently simple to admit an exact solution, thereby making transparent the mathematical derivations presented in [1]. The general theory underlying *SO-ASAP* indicates that, for a physical system comprising $N_\alpha$ parameters, the computation of all of the first- and second-order response sensitivities requires, in principle, $(2N_\alpha + 1)$ "large-scale" computations involving correspondingly constructed adjoint systems, which we called *second adjoint sensitivity systems (SASS)*. However, the illustrative application presented in this work has shown that the actual number of adjoint computations needed for computing all of the first- and second-order response sensitivities may be far less than $(2N_\alpha + 1)$ per response. In particular for this illustrative problem, four (4) "large-scale" adjoint computations sufficed for the complete and exact computations of all 4 first- and 10 distinct second-order derivatives. We have also shown that the construction and solution of the *second adjoint sensitivity system* (*SASS*) requires very little additional effort beyond the construction of the adjoint sensitivity system needed for computing the first-order sensitivities. Very significantly, only the sources on the right-sides of the solver for the diffusion equation needed to be modified, while the left-side of the differential equations remained unchanged.

All of the first-order response sensitivities to the model parameters had significant values, i.e., relative values of (or approaching) unity. After discussing the physical significance of the first-order sensitivities, we showed that most of the second-order relative sensitivities were comparable in magnitude (i.e., as large as, or larger than) to that of the first-order relative sensitivities. We have subsequently showed that the second-order sensitivities play the following important roles:
  (a) they cause the "expected value of the response" to differ from the "computed nominal value of the response";
  (b) they contribute to increasing the response variances and modifying the response covariances, but their contribution is smaller than that stemming from the first-order response sensitivities; and
  (c) they *contribute decisively to causing assymetries in the response distribution*.

Indeed, neglecting the second-order sensitivities would nullify the third-order response correlations, and hence would nullify the *skewness* of the response. Consequently, any events occurring in a response's long and/or short tails, which are characteristic of rare but decisive events (e.g., major accidents, catastrophies), would likely be missed.

We are currently in the process of documenting the generalization of the *SO-ASAP*, turning into a *high-order adjoint sensitivity analysis procedure* (*HO-ASAP*), which will enable the exact computation of arbitrarily high-order response sensitivities in an efficient manner. The *HO-ASAP* will enable the *exact* computation of all of the $K^{th}$-order response derivatives to $N$ model parameters *in at most $O(N^{K-1})$ computations*. This innovation is expected to impact significantly the fields of optimization and predictive modeling, including data assimilation/adjustment and model calibration.

**ACKNOWLEDGMENTS**

This work was partially supported by contract 15540-FC51 from Gen4Energy, Inc., and partially by contract 15540-FC59 from the US Department of Energy (NA-22), respectively, with the University of South Carolina. The author wishes to express his personal appreciation to Mr. Regan Voit, Gen4Energy VP for Applied Research and Manufacturing Initiatives, and Mr. James Peltz, NA-22 Program Manager, for their continued support.